\documentclass[11pt]{article}
\usepackage{cite}      
\usepackage{graphicx}  
\usepackage{indentfirst}
\usepackage{amsmath}
\usepackage{amssymb}
\usepackage{amsthm}
\usepackage{amsfonts}

\usepackage{anysize}
\usepackage{fancyhdr}
\title{Relay Subset Selection in Wireless Networks Using Partial Decode-and-Forward Transmission}
\author{Caleb K. Lo, Sriram Vishwanath and Robert W. Heath, Jr. \\Wireless Networking and Communications Group \\ Department of Electrical and Computer Engineering \\ The University of Texas at Austin \\ 1 University Station C0803 \\ Austin, TX 78712-0240 \\ Phone: (512) 471-1190 \\ Fax: (512) 471-6512 \\ Email: \{clo, sriram, rheath\}@ece.utexas.edu}
\date{}

\begin{document}

\maketitle

\begin{abstract}
This paper considers the problem of selecting a subset of nodes in a two-hop wireless network to act as relays in aiding the communication between the source-destination pair.  Optimal relay subset selection with the objective of maximizing the overall throughput is a difficult problem that depends on multiple factors including node locations, queue lengths and power consumption.  A partial decode-and-forward strategy is applied in this paper to improve the tractability of the relay selection problem and performance of the overall network.

Note that the number of relays selected ultimately determines the performance of the network.  This paper benchmarks this performance by determining the net diversity achieved using the relays selected and the partial decode-and-forward strategy.  This framework is subsequently used to further transform relay selection into a simpler relay placement problem, and two proximity-based approximation algorithms are developed to determine the appropriate set of relays to be selected in the network.  Other selection strategies such as random relay selection and a greedy algorithm that relies on channel state information are also presented.  This paper concludes by showing that the proposed proximity-based relay selection strategies yield near-optimal expected rates for a small number of selected relays.
\end{abstract}

Keywords - Greedy algorithms, partial decode-and-forward, superposition coding, relays.

\section{Introduction}
Relay-assisted communication is a promising strategy for both centralized and decentralized communication networks \cite{ESSMeshNetw, RelaTaskGrou}.  Two-hop relay-based communication is having a considerable influence on emerging standards both in local area networks, IEEE 802.11s \cite{ESSMeshNetw} and broadband wireless access networks, IEEE 802.16j \cite{RelaTaskGrou}.  Two-hop relay systems consist of a source, a destination and one or more relays where the relay nodes work together as a single set of intermediaries between the source and the destination \cite{PabETAL:RelaBaseDeplConc:Sep:04}.  Direct transmission occurs between the source and the destination, and the relays assist the source only if the destination cannot decode the direct transmission.  There are multiple concrete benefits of introducing these intermediate relays, which include improved system throughput and greater coverage \cite{RelaTaskGrou}.  Multihop relaying \cite{BoyETAL:MultDiveWireRela:Oct:04, DohETAL:CapaDistPhyLaye:Mar:06} is a key enabling technology for networks of the future, but before the performance tradeoffs of multihop relaying can be characterized, it is critical that the issues facing two-hop relaying be fully understood.

Given that the source can enlist multiple nodes to simultaneously act as relays, two questions naturally arise.  First, how many relays must the source enlist to aid its transmission to gain the maximum advantage for the resources consumed? Second, which of the nodes in the pre-existing network must be enlisted to act as relays?  When multiple-relay selection is allowed, there are numerous tradeoffs that govern system performance \cite{LanWor:DistSpacTimeCode:Oct:03, DohETAL:CapaDistPhyLaye:Mar:06, JinHas:DistSpacTimeCodi:Dec:06}.  While selecting a large number of relays offers the benefit of coherent combining, resulting in increased throughput and thus higher overall quality of service, it suffers from drawbacks as well. Firstly, system resources are drained faster when multiple relays are selected.  Second, there are complexity and implementation issues - it is difficult to synchronize the transmissions from multiple disparate relays \cite{HuSer:ScalCoopTimeSync:Jun:06, WeiETAL:AsynCoopDive:Jun:06, SunETAL:SecuResiClocSync:Feb:06, LinETAL:NetwSync:Oct:85}, and receiver complexity increases with the number of relays.  A single relay can be selected to assist the source transmission \cite{ZhaTod:CellCDMACapaOut:Feb:06, TarMinETAL:DiffModuTwoUser:Sep:05, VenWanETAL:MultDeteCoopNetw:Sep:06, ZorRao:GeogRandForwAd:Oct:03, BleKhiETAL:SimpCoopDiveMeth:Mar:06, LiuSu:OptiSeleRelaProt:Nov:06, GunErk:OppoCoopDynaReso:Apr:07, ZhaVal:PracRelaNetwGene:Jan:05}, which offers lower gains in terms of total diversity and rate but is simpler to implement and consumes less power over the entire network.  This paper has two goals.  One goal is to understand the fundamental limits of multiple-relay selection to benchmark various relay selection algorithms.  To this end, we focus on minimizing relay power consumption and treat implementation issues and complexity as a secondary concern.

Regardless of the number of relays selected, it is difficult to determine which node(s) in the network must act as relays to aid the source transmission.  For example, selecting the relay with the best channel to the destination may not be an optimal strategy, as this relay may be heavily loaded with traffic and running low on resources.  Thus, relay selection is a very difficult problem, as selecting the ``optimal'' subset from the set of candidate relay nodes is affected by the presence of multiple parameters that govern system performance.  In particular, relay node selection often translates to a combinatorial optimization problem \cite{CooCunETAL:CombOpti:97}, which currently does not have an elegant polynomial-time algorithmic solution.

The second goal of this paper is to provide algorithms for relay node selection that serve as a good approximation to the problem of optimal relay selection from the point of view of throughput maximization with power allocation.  Moreover, we desire the algorithms to have low complexity and be highly intuitive in terms of design.  Note that any selection algorithm is closely coupled with the transmission strategy employed in the network (such as decode/amplify/compress-and-forward).  Thus, we discuss the transmission strategy employed in this paper and then delve into the details of the algorithms.

In our paper, we use a partial decode-and-forward transmission strategy proposed in \cite{YukErk:BroaStraFadiRela:Oct:04}\footnote{Note that this notion of partial decode-and-forward is distinct from the one in \cite{CovGam:CapaTheoRelaChan:Sep:79} as it is inspired by outage capacity.  It is based on the superposition coding strategy for broadcast channels in \cite{Sha:BroaStratGausSlow:Jun:97}, while the partial decode-and-forward strategy in \cite{CovGam:CapaTheoRelaChan:Sep:79} is derived from block-Markov coding.}.  Partial decode-and-forward as described in \cite{YukErk:BroaStraFadiRela:Oct:04} relies on a two-level superposition coding strategy introduced by T. Cover for broadcast channels \cite{Cov:BroaChan:Jan:72} and further studied in \cite{ShaSte:BroaApprSingUser:Oct:03, LiuLauETAL:OptiRateAlloSupe:Jun:02, RocYeuETAL:SymmMultDiveCodi:May:97}.  Under this setting, the transmitter employs a layered coding strategy, allowing the receiver to decode the transmitter's message partially if it is incapable of determining it in its entirety.  Note that the conventional decode-and-forward strategy as in \cite{CovGam:CapaTheoRelaChan:Sep:79} is a special case of the partial decode-and-forward strategy, and therefore, partial decode-and-forward is a useful tool that has all the properties of decode and forward incorporated into it.  In particular, partial decode-and-forward offers both the diversity advantages of amplify-and-forward and the inherent robustness to noise of decode-and-forward \cite{LanTseETAL:CoopDiveWireNetw:Dec:04}.

The other main advantage of partial decode-and-forward is the tractability it lends to the relay selection problem.  While multiple-relay selection based on partial decode-and-forward transmission does not readily lend itself to practical implementation, the resulting problem tractability facilitates the determination of valuable performance benchmarks, especially in terms of diversity gain.  Our first contribution is the derivation of both the diversity gain and the generalized diversity gain that is achieved by allowing $m$ of the candidate relays to assist the source.  The resulting diversity analysis extends the single-relay result in \cite{YukErk:BroaStraFadiRela:Oct:04} and further highlights the performance benefits of multiple-relay selection.  We stress that the derived diversity gain is at most the diversity achieved by selecting $m$ relays out of $K_r$ candidate relays.  For example, selecting the relay with the best end-to-end path between the source and the destination can yield a diversity gain of $K_r+1$ \cite{BleKhiETAL:SimpCoopDiveMeth:Mar:06}.

We mention here that generalized diversity, which arises from the notion of generalized degrees of freedom \cite{JafVis:GeneDegrFreeSymm, EtkTseETAL:GausInteChanCapa:Feb:07}, refers to the diversity achieved when the candidate relays have different transmit SNR values than that of the source.  In our paper, we consider a specific case of generalized diversity where the SNR for each candidate relay is an exponential scaling of the source SNR, i.e. $(P_i/\sigma^2) = (P_t/\sigma^2)^k$ for relay $i$.  This allows for performance benchmarking of networks where the relays may be operating on a different power budget than that of the source, including relay-assisted cellular networks.  For example, the case where $k < 1$ can be modeled by a base station being assisted by mobile devices that are not currently handling their own voice traffic.  On the other hand, the case where $k > 1$ can be modeled by a battery-powered mobile device being assisted by fixed, dedicated relay nodes that are connected to a continuous power source.

Our second contribution entails using the partial decode-and-forward framework as a platform to transform the relay selection problem into a relay-placement problem, whose solution suggests the ``best'' set of relays to be selected.  There are two approximation steps: we first approximate selection of $m$ relays by the problem of finding the $m$ relays that are closest to a rate-maximizing location, and we show that obtaining the rate-maximizing location is equivalent to maximizing a $\textit{signomial function}$ \cite{BoyVan:ConvOpti:04}.  Since $\textit{signomial programs}$ are, in general, not easy to solve, we further approximate the relay selection problem by the problem of finding the relays that are closest to the rate-maximizing location in a three-node line network.  Obtaining this rate-maximizing location is equivalent to maximizing a polynomial over a given range of values, which can be accomplished using deterministic polynomial-time algorithms.  The above polynomial approximation motivates two $\textit{proximity-based}$ algorithms (which we call Multiple Fan Out and Single Fan Out, detailed in Section \ref{subset-sel-alg}) that select relays based on their proximity to one of the rate-maximizing locations.  In addition, we present a greedy selection algorithm (which we call Best Gains, also detailed in Section \ref{subset-sel-alg}) that chooses relays based on their channel gains to the destination and the amount of the source message that they have decoded.  Here, the selected relays must have decoded at least one of the two messages from the source.  Finally, we present a selection algorithm that randomly selects relay nodes (which we call Random Relays, also detailed in Section \ref{subset-sel-alg}) and compare the performance of all four algorithms - Multiple Fan Out, Single Fan Out, Best Gains and Random Relays.


This paper is organized as follows.  In Section II we describe the system model and introduce the two-level superposition coding strategy that will be used throughput the paper.  The diversity analysis is shown in Section III.  In Section IV, we present the analytical formulation of the relay selection problem and obtain a closed-form expression for the rate-maximizing relay position in a three-node line network.  We present our proposed selection algorithms in Section V.  We present simulation results in Section VI and conclude the paper in Section VII.

\section{System Model}
First, we introduce the notation used throughout the paper.  $\mathbb{E}$ denotes the mathematical expectation operator and $\ln(\cdot)$ represents the natural logarithm function.  $\exp(\cdot)$ represents the exponential function and $\Gamma(\cdot)$ is the gamma function.  SNR represents the transmit-side signal-to-noise ratio at the source node.  $P(A)$ denotes the probability that an event $A$ occurs.  $f^{'}(x)$ denotes the derivative of a function $f$ with respect to its argument $x$.  $\|\mathcal{A}\|$ denotes the cardinality of a set $\mathcal{A}$.  $|z|^2$ denotes the absolute square of a complex number $z$.  $f(x)\sim g(x)$ for large values of $x$ represents the fact that $f(x)/g(x)\rightarrow 1$ as $x\rightarrow\infty$ \cite[Pg. 3067]{LanTseETAL:CoopDiveWireNetw:Dec:04}.

Consider the two-hop wireless network in Fig. \ref{system-model}.  The network consists of a single source $t$, a single destination $r$ and $K_r$ relays interspersed throughout the region between the source and the destination.  Let $d_{i,n}$ denote the distance between nodes $i$ and $n$.  Let $h_{i,n}$ denote the channel between nodes $i$ and $n$.  

\subsection{Key Assumptions}
We make the following critical assumptions in this paper:
\begin{itemize}
\item Each relay operates in a half-duplex mode and employs a single antenna.
\item Additive white circularly symmetric complex Gaussian noise $n_{i,k}$ with mean 0 and variance $\sigma^2$ is present at each receiving node $i$ during time slot $k$.
\item $|h_{i,n}|$ is a Rayleigh-distributed random variable.  Thus, the real and imaginary components of $h_{i,n}$ are mutually independent Gaussian-distributed random variables, each with mean 0 and variance $(1/2)\cdot\mathbb{E}(|h_{i,n}|^2)$.  This assumption simplifies our analysis and is typically used in the literature to obtain insights on the performance of real-world wireless systems.
\item Our transmission strategy has arbitrarily long codewords that are generated using an i.i.d. Gaussian distribution of suitable variance to meet the overall power constraint.  Note that the capacity and thus, the capacity achieving input distribution (if any) of the additive Gaussian noise relay channel is in general unknown.  Thus, we choose this coding strategy for two reasons: 1) it has been found to be optimal in most of the special cases whose capacity is known (physically/reversely degraded \cite{CovGam:CapaTheoRelaChan:Sep:79}, orthogonal channels \cite{GamZah:CapaClasRelaChan:May:05} and uniform phase fading \cite{KraETAL:CoopStraCapaTheo:Sep:05}) and 2) it yields explicit rate expressions and intuitive coding strategies.  The destination and all potential relay nodes employ typical set decoding as defined in \cite{CovTho:ElemInfoTheo:06}.
\item The source knows the exact channel state for all of the channels in the network in Fig. \ref{system-model}.  Each relay knows the exact state of its channel from the source.  The destination knows the exact state of its channel from each of the relays and from the source.
\item Time is slotted and the channel is constant in every time slot.
\item Each time slot is large enough to admit an arbitrarily small probability of error as long as the rate in that state is below the maximum achievable for that state (this is also referred to as the block fading assumption).
\item A log-distance path loss model is applied \cite{Rap:WireCommPrinPrac:02}.  Let $\lambda_c$, $d_0$, and $\mu$ denote the carrier wavelength, the reference distance, and the path loss exponent.  Then, the channel gain between nodes $i$ and $n$ is 
\begin{eqnarray}
\mathbb{E}(|h_{i,n}|^2) & = & G_{i,n}^2 \nonumber \\
& = & (\lambda_c/4\pi d_0)^2(d_{i,n}/d_0)^{-\mu}. \label{path-loss}
\end{eqnarray}  
\end{itemize}

\subsection{Partial Decode-and-Forward}
All relays perform partial decode-and-forward operations based on the two-level superposition coding strategy in \cite{YukErk:BroaStraFadiRela:Oct:04}.  The source transmits $x_{t,1}$ during the first time slot, where
\begin{equation}
x_{t,1} = x_1 + x_2
\end{equation}
and the source allocates power $\beta P_t$ to $x_1$ and power $\bar{\beta}P_t$ to $x_2$, where $\beta\in [0,1]$ and $\bar{\beta} = 1-\beta$.  Note that $x_1$ and $x_2$ are codewords from codebooks with elements that are generated i.i.d. according to zero-mean Gaussian distributions with variance $\beta P_t$ and $\bar{\beta}P_t$, respectively.

The destination and all candidate relay nodes employ typical set decoding to decode $x_1$ and $x_2$.  The candidate relays and the destination initially attempt to decode $x_1$.  If node $i$ can decode $x_1$ then it attempts to decode $x_2$.  Two channel thresholds, $|h_1|$ and $|h_2|$, are chosen to determine the set of received rates for this two-level coding strategy.  Then, $x_1$ can be decoded at the rate $R_1$ \cite{YukErk:BroaStraFadiRela:Oct:04}, where
\begin{equation}\label{rate-x1}
R_1 = \ln\bigg(1+\frac{|h_1|^2\beta P_t}{|h_1|^2\bar{\beta}P_t+\sigma^2}\bigg)
\end{equation}
while $x_2$ can be decoded at the rate $R_2$ \cite{YukErk:BroaStraFadiRela:Oct:04}, where
\begin{equation}\label{rate-x2}
R_2 = \ln\bigg(1+\frac{|h_2|^2\bar{\beta}P_t}{\sigma^2}\bigg).
\end{equation}
Note that if node $i$ attempts to decode $x_1$ or $x_2$ at a higher rate than $R_1$ or $R_2$, respectively, the resulting probability of error is bounded away from zero.

The received signals at the candidate relay $i$ and at the destination during time slot 1 are, respectively
\begin{eqnarray}
y_{i,1} & = & h_{t,i}x_{t,1}+n_{i,1} \\
y_{r,1} & = & h_{t,r}x_{t,1}+n_{r,1}.
\end{eqnarray}
If the destination can decode both $x_1$ and $x_2$, it broadcasts this information to the entire network and the source prepares to send $x_{t,2}$ during time slot 2.  If the destination can only decode $x_1$, or if it cannot decode $x_1$, it broadcasts this information to the entire network.  The source then selects a subset of the candidate relays to assist its transmission.

For relay $i$, if $|h_{t,i}| < |h_1|$, then it cannot decode $x_1$ and it does not transmit during time slot 2.  If $|h_1| \leq |h_{t,i}| < |h_2|$, then a selected relay $i$ can only decode $x_1$ and will forward $x_1$ to the destination during time slot 2.  If $|h_{t,i}| \geq |h_2|$, then a selected relay $i$ can decode $x_{t,1}$ and will forward $x_{t,1}$ to the destination.  Thus, relay $i$ allocates power $P_i$ to its transmission $x_{r,i}$ to the destination, where \cite{YukErk:BroaStraFadiRela:Oct:04}
\begin{equation}
x_{r,i} = \left\{\begin{array}{ll}
0 & \textnormal{if } |h_{t,i}| < |h_1| \\
\sqrt{\frac{P_i}{\beta P_t}}x_1 & \textnormal{if } |h_1| \leq |h_{t,i}| < |h_2| \\
\sqrt{\frac{P_i}{P_t}}(x_1+x_2) & \textnormal{if } |h_{t,i}| \geq |h_2|.
\end{array}\right.
\end{equation}
For each relay $i$, we set $\beta_i = \beta$ for the majority of this paper; in Section \ref{perf-rel-sel-alg} we investigate the performance impact of varying $\beta_i$ with respect to $\beta$.

Thus, if $\mathcal{A}$ denotes the set of all relays that transmit during time slot 2, the destination receives
\begin{equation}
y_{r,2} = \sum_{i \in \mathcal{A}}h_{i,r}x_{r,i}+n_{r,2}
\end{equation}
during time slot 2.  After time slot 2, if the destination can decode $x_{t,1}$, the received rate is $R_1 + R_2$.  If the destination can only decode $x_1$, the received rate is $R_1$, and if the destination cannot decode $x_1$, the received rate is 0.  Note that this two-level coding strategy can be generalized to a multiple-level approach based on broadcast strategies introduced for the single user and MAC channels \cite{Sha:BroaStratGausSlow:Jun:97}.  Once the two-level strategies and algorithms are understood, their generalization to $n>2$ levels is relatively straightforward but leads to extremely unwieldy expressions.  Moreover, it is unclear if using a multiple-level approach will provide significant gains in performance.  Thus, we have chosen to limit ourselves to a two-level transmission strategy in this paper.

Our proposed multiple-relay selection algorithms choose $\mathcal{A}$ to maximize the expected rate subject to a sum power constraint over all relays $i\in\mathcal{A}$.  In the next section we derive both the diversity gain and the generalized diversity gain via selecting $m$ relays.

\section{Diversity Performance}\label{pwr-alloc-strat}
After $m$ relays are selected to transmit to the destination during time slot 2, we consider the resulting diversity gains $\kappa_1(m)$ and $\kappa_2(m)$ for $R_1$ and $R_2$, respectively.  Let $P_{out}(R_1,\mathcal{A})$ denote the probability that the destination cannot decode $x_1$ after time slot 2, and let $P_{out}(R_2,\mathcal{A})$ denote the probability that the destination cannot decode $x_2$ after time slot 2.

For the diversity analysis, we set the relay powers $P_i = P_t$ for $i\in\{1,2,\ldots,m\}$, and so the SNR is $P_t/\sigma^2$.  The diversity gains are obtained by observing that the outage probabilities $P_{out}(R_1,\mathcal{A})$ and $P_{out}(R_2,\mathcal{A})$ are proportional to $SNR^{-\kappa_1(m)}$ and $SNR^{-\kappa_2(m)}$, respectively as the SNR $P_t/\sigma^2$ approaches infinity.  We reiterate that these diversity gains are at most the diversity achieved by selecting $m$ relays out of $K_r$ candidate relays.

\newtheorem{m-relays}{Theorem}
\begin{m-relays}\label{m-relays}
Selecting $m$ relays to transmit during time slot 2 yields a diversity gain of
\begin{eqnarray}
\kappa_1(m) & = & m+1 \nonumber
\end{eqnarray}
and $\kappa_2(m) = \kappa_1(m)$.
\end{m-relays}

\begin{proof}
See Appendix \ref{proof-thm-1}.
\end{proof}

We note that obtaining the diversity gain of $m+1$ entails a relatively straightforward extension of the single-relay analysis for decode-and-forward relaying in \cite[Section IV.B]{LanTseETAL:CoopDiveWireNetw:Dec:04}.  In the two-level transmission strategy that we consider, the destination still attempts to decode the transmission from the source even if at least one relay fails to decode $x_1$ or $x_2$.  Note that the single-relay analysis in \cite[Section IV.B]{LanTseETAL:CoopDiveWireNetw:Dec:04} ignores the direct link transmission if the relay makes a decoding error, and so a direct extension of the analysis in \cite[Section IV.B]{LanTseETAL:CoopDiveWireNetw:Dec:04} would yield a diversity gain of $m$ instead.

We also perform a generalized diversity analysis where the relay powers are such that $(P_i/\sigma^2) = (P_t/\sigma^2)^k$ for $i\in\{1,2,\ldots,m\}$, where $k$ is a real number.  The generalized diversity gains $\kappa_1^g(m)$ and $\kappa_2^g(m)$ are obtained by observing that the outage probabilities $P_{out}(R_1,\mathcal{A})$ and $P_{out}(R_2,\mathcal{A})$ are proportional to $SNR^{-\kappa_1^g(m)}$ and $SNR^{-\kappa_2^g(m)}$, respectively as the transmit-side SNR $P_t/\sigma^2$ approaches infinity.

\newtheorem{m-relays-general}[m-relays]{Theorem}
\begin{m-relays-general}\label{m-relays-general}
Selecting $m$ relays to transmit during time slot 2 yields a generalized diversity gain of
\begin{eqnarray}
\kappa_1^g(m) & = & \left\{ \begin{array}{ll}
km+1 & \textnormal{if } k\leq 1 \nonumber \\
m+1 & \textnormal{if } k > 1
\end{array} \right.
\end{eqnarray}
and $\kappa_2^g(m) = \kappa_2^g(m)$.
\end{m-relays-general}

\begin{proof}
See Appendix \ref{proof-thm-2}.
\end{proof}

The generalized diversity gain $\kappa_1^g(m) = \kappa_2^g(m)$ has the following intuitive interpretation.  If $k\leq 1$, each relay is no better than the source in terms of transmit power, so the worst-case error event is determined by all of the relay-to-destination channels.  In particular, this event occurs when all $m$ suboptimal relays attempt to forward $x_1$ or $x_2$ to the destination.  On the other hand, if $k > 1$, each relay is better than the source in terms of transmit power, so the worst-case error event is determined by all of the source-to-relay channels.  In particular, this event occurs when all $m$ superior relays cannot decode either $x_1$ or $x_2$.

\section{Rate-Maximizing Relay Position}\label{opt-relay-place}
We formulate the relay selection problem for an arbitrary number of selected relays, and then we show how this problem can be simplified by considering a three-node line network.

\subsection{Optimal Relay Placement in General Network}\label{m-nodes}
Consider the case where a subset $\mathcal{A}$ of the available relay nodes $\{1,2,\ldots,K_r\}$ are selected to assist the source.  Let $h$ denote the channel between a transmitting node and a receiving node.  The received rate at a receiving node via decoding $x_1$ is \cite{YukErk:BroaStraFadiRela:Oct:04}
\begin{equation}
C_1(|h|^2) \triangleq \ln\bigg(1+\frac{|h|^2\beta P_t}{|h|^2\bar{\beta}P_t+\sigma^2}\bigg)
\end{equation}
and the received rate at a receiving node via decoding $x_2$ after decoding $x_1$ is \cite{YukErk:BroaStraFadiRela:Oct:04}
\begin{equation}
C_2(|h|^2) \triangleq \ln\bigg(1+\frac{|h|^2\bar{\beta}P_t}{\sigma^2}\bigg).
\end{equation}

The expected rate of the two-level superposition coding strategy is
\begin{equation}\label{r-sc-m}
\bar{R}_{sc,2}(\mathcal{A}) = (1-P_{out}(R_1,\mathcal{A}))R_1 + (1-P_{out}(R_1,\mathcal{A}))(1-P_{out}(R_2,\mathcal{A}))R_2
\end{equation}
and so the relay selection problem can be formulated as follows
\begin{eqnarray}
& & \max_{\mathcal{A}\subseteq\{1,2,\ldots,K_r\}} \bar{R}_{sc,2}(\mathcal{A}) \label{signom-func} \\
& & \textnormal{subject to } \sum_{i\in\mathcal{A}} P_i \leq P_{max}\textnormal{ and }0\leq P_i\leq P_{i,max}\quad\forall i\in\mathcal{A}. \nonumber
\end{eqnarray}
It is apparent from \eqref{signom-func} that the relay selection problem is also a power allocation problem.  In particular, if a relay $i$ is not selected, it is assigned a power $P_i = 0$.  On the other hand, if a relay $i$ is selected, it is assigned a power $P_i > 0$ according to the solution to \eqref{signom-func}.

Let $\Delta$ denote the set of all relays that cannot decode $x_1$, and let $\Theta$ denote the set of all relays that can decode $x_1$ but cannot decode $x_2$.  The probability that the destination cannot decode $x_1$ after time slot 2 can be obtained by generalizing \cite[(13)]{YukErk:BroaStraFadiRela:Oct:04} as
\begin{eqnarray}
P_{out}(R_1,\mathcal{A}) & = & \sum_{(0\leq\alpha,\xi\leq\|\mathcal{A}\|),\alpha+\xi\leq\|\mathcal{A}\|}\Bigg(\sum_{\Delta\subseteq\mathcal{A}, \Theta\subseteq\mathcal{A}, \|\Delta\| = \alpha, \|\Theta\| = \xi, \Delta\bigcap\Theta = \emptyset}\Bigg(\Bigg(\prod_{\delta\in\Delta}P(C_1(|h_{t,\delta}|^2) < R_1)\Bigg) \nonumber \\
& & \times\Bigg(\prod_{\theta\in\Theta}P(C_1(|h_{t,\theta}|^2) \geq R_1, C_2(|h_{t,\theta}|^2) < R_2)\Bigg)\Bigg(\prod_{\eta\in (\mathcal{A}\setminus (\Delta\bigcup\Theta))}P(C_2(|h_{t,\eta}|^2) \geq R_2)\Bigg)\Bigg) \nonumber \\
& & \times P\Bigg(\ln\Bigg(1+\frac{|h_{t,r}|^2\beta P_t+\sum_{\eta\in (\mathcal{A}\setminus (\Delta\bigcup\Theta))}|h_{\eta,r}|^2\beta P_\eta}{|h_{t,r}|^2\bar{\beta}P_t+\sum_{\eta\in (\mathcal{A}\setminus (\Delta\bigcup\Theta))}|h_{\eta,r}|^2\bar{\beta} P_\eta+\sigma^2}+\sum_{\theta\in\Theta}\frac{|h_{\theta,r}|^2P_\theta}{\sigma^2}\Bigg) < R_1\Bigg)\Bigg). \label{p-out-R1-m}
\end{eqnarray}
Each term in the inner sum in \eqref{p-out-R1-m} represents a scenario where $\alpha$ selected relays cannot decode $x_1$, $\xi$ selected relays can decode $x_1$ but cannot decode $x_2$, and the remaining $\|\mathcal{A}\|-\alpha-\xi$ selected relays can decode $x_2$.

The expressions in \eqref{p-out-R1-m} are fairly involved, so we consider the high-SNR regime for ease of analysis.  In Appendix \ref{proof-thm-1}, we prove that
\begin{eqnarray}
P(C_1(|h_{t,\delta}|^2) < R_1) & \sim & \frac{1}{G_{t,\delta}^2}\times\frac{\exp(R_1)-1}{(P_t/\sigma^2)\times (1-\bar{\beta}\exp(R_1))} \label{p-out-R1-m-1} \\
P(C_1(|h_{t,\theta}|^2) \geq R_1, C_2(|h_{t,\theta}|^2) < R_2) & \sim & 1 \label{p-out-R1-m-2} \\
P(C_2(|h_{t,\eta}|^2) \geq R_2) & \sim & 1 \label{p-out-R1-m-3}
\end{eqnarray}
and
\begin{eqnarray}
P\Bigg(\ln\Bigg(1+\frac{|h_{t,r}|^2\beta P_t+\sum_{\eta\in (\mathcal{A}\setminus (\Delta\bigcup\Theta))}|h_{\eta,r}|^2\beta P_\eta}{|h_{t,r}|^2\bar{\beta}P_t+\sum_{\eta\in (\mathcal{A}\setminus (\Delta\bigcup\Theta))}|h_{\eta,r}|^2\bar{\beta} P_\eta+\sigma^2}+\sum_{\theta\in\Theta}\frac{|h_{\theta,r}|^2P_\theta}{\sigma^2}\Bigg) < R_1\Bigg) & & \nonumber \\
\sim \Bigg(\frac{\exp(R_1)-1}{(P_t/\sigma^2)\times (1-\bar{\beta}\exp(R_1))}\Bigg)^{\|\mathcal{A}\|-\alpha+1}\times\frac{1}{(\|\mathcal{A}\|-\alpha+1)!}\times\frac{1}{G_{t,r}^2}\prod_{\nu\in (\mathcal{A}\setminus\Delta)}\frac{1}{(P_{\nu}/P_t)G_{\nu,r}^2}. & & \label{p-out-R1-m-4}
\end{eqnarray}

The probability that the destination cannot decode $x_2$ after time slot 2 is
\begin{eqnarray}
P_{out}(R_2,\mathcal{A}) & = & \sum_{0\leq\alpha\leq\|\mathcal{A}\|}\Bigg(\sum_{\Delta\subseteq\mathcal{A}, \|\Delta\| = \alpha}\Bigg(\Bigg(\prod_{\delta\in\Delta}P(C_2(|h_{t,\delta}|^2) < R_2)\Bigg) \nonumber \\
& & \times\Bigg(\prod_{\theta\in (\mathcal{A}\setminus\Delta)}P(C_2(|h_{t,\theta}|^2) \geq R_2)\Bigg) \nonumber \\
& & \times P\Bigg(C_2\Bigg(|h_{t,r}|^2+\sum_{\theta\in (\mathcal{A}\setminus\Delta)}|h_{\theta,r}|^2\Bigg) < R_2\Bigg)\Bigg)\Bigg). \label{p-out-R2-m}
\end{eqnarray}
Each term in the inner sum in \eqref{p-out-R2-m} represents a scenario where $\alpha$ selected relays cannot decode $x_2$ and the remaining $\|\mathcal{A}\|-\alpha$ selected relays can decode $x_2$.

The expressions in \eqref{p-out-R2-m} are also fairly involved, so we again consider the high-SNR regime for ease of analysis.  In Appendix \ref{proof-thm-1}, we prove that
\begin{eqnarray}
P(C_2(|h_{t,\delta}|^2) < R_2) & \sim & \frac{1}{G_{t,\delta}^2}\times\frac{\exp(R_2)-1}{\bar{\beta}(P_t/\sigma^2)} \label{p-out-R2-m-1} \\
P(C_2(|h_{t,\theta}|^2) \geq R_2) & \sim & 1 \label{p-out-R2-m-2}
\end{eqnarray}
and
\begin{eqnarray}
P\Bigg(C_2\Bigg(|h_{t,r}|^2+\sum_{\theta\in (\mathcal{A}\setminus\Delta)}|h_{\theta,r}|^2\Bigg) < R_2\Bigg) & & \nonumber \\
\sim \Bigg(\frac{\exp(R_2)-1}{\bar{\beta}(P_t/\sigma^2)}\Bigg)^{\|\mathcal{A}\|-\alpha+1}\times\frac{1}{(\|\mathcal{A}\|-\alpha+1)!}\times\frac{1}{G_{t,r}^2}\prod_{\theta\in (\mathcal{A}\setminus\Delta)}\frac{1}{(P_{\theta}/P_t)G_{\theta,r}^2}. & & \label{p-out-R2-m-3}
\end{eqnarray}

It is apparent that \eqref{signom-func} is an optimization problem with linear inequality constraints.  Also, from inspecting \eqref{p-out-R1-m}-\eqref{p-out-R1-m-4} it is clear that $P_{out}(R_1,\mathcal{A})$ is a nonlinear function of $P_i~\forall~i\in\mathcal{A}$ in the high-SNR regime.  In addition, from inspecting \eqref{p-out-R2-m}-\eqref{p-out-R2-m-3} it is clear that $P_{out}(R_2,\mathcal{A})$ is a nonlinear function of $P_i~\forall~i\in\mathcal{A}$ in the high-SNR regime.  Then, the preceding analysis shows that in the high-SNR regime, $\bar{R}_{sc,2}(\mathcal{A})$ is a nonlinear function of $P_i$ for $i\in\mathcal{A}$.  Thus, nonlinear programming techniques such as sequential quadratic programming \cite{NasSof:LineNonlProg:96} can be applied to solve \eqref{signom-func} in the high-SNR regime.

\begin{figure}[tb]
\begin{center}
\includegraphics[width=3.0in]{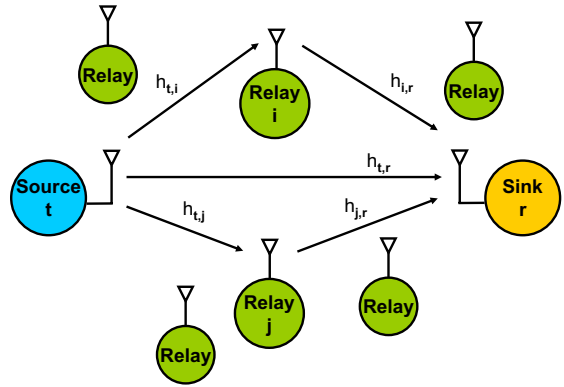}
\end{center}
\caption{Two-hop wireless network.}
\label{system-model}
\end{figure}

The relay selection problem \eqref{signom-func} can also be approximated as a relay placement problem where $m$ relays in Fig. \ref{system-model} are chosen to assist the source.  The key idea behind the relay placement problem is to hypothetically place $m$ relays in the locations that would maximize $\bar{R}_{sc,2}(\mathcal{A})$.  Then, the $m$ relays in Fig. \ref{system-model} that are closest to the rate-maximizing locations are selected.  It is also assumed that each selected relay $i$ employs the same power $P_i = P_{max}/m$.  

To solve for the rate-maximizing locations, recall from \eqref{path-loss} that $G_{i,n}^2 = (\lambda_c/4\pi d_0)^2(d_{i,n}/d_0)^{-\mu} = (d_{i,n})^{-\mu}\chi$.  Without loss of generality, assume that the source is located at (0,0) and the destination is located at $(d_{t,r},0)$.  If relay $i$ is located at $(a_i,b_i)$, then $d_{t,i} = \sqrt{a_i^2+b_i^2}$ and $d_{i,r} = \sqrt{(d_{t,r}-a_i)^2+b_i^2}$.  The rate-maximizing locations are
\begin{eqnarray}
\{a_1^{*},b_1^{*},\ldots,a_m^{*},b_m^{*}\}& = & \textnormal{arg}\max_{a_1,b_1,\ldots,a_m,b_m} \bar{R}_{sc,2}(\mathcal{A}) \nonumber \\
& & \textnormal{subject to } \|\mathcal{A}\| = m, \sum_{i\in\mathcal{A}} P_i \leq P_{max}\textnormal{ and }0\leq P_i\leq P_{i,max}\quad\forall i\in\mathcal{A}. \nonumber
\end{eqnarray}

In particular, by considering the binomial series $\sum_{k=0}^{\infty}(a+b)^k$ where $k$ is a real number, we see that
\begin{eqnarray}
d_{t,i}^{\mu} & = & (a_i^2+b_i^2)^{\mu/2} \nonumber \\
& = & \sum_{k=0}^{\infty}\frac{\Gamma(\mu/2+1)}{k!\Gamma(\mu/2+1-k)}a_i^{2k}b_i^{\mu-2k} \label{d-t-i}
\end{eqnarray}
and
\begin{eqnarray}
d_{i,r}^{\mu} & = & ((d_{t,r}-a_i)^2+b_i^2)^{\mu/2} \nonumber \\
& = & \sum_{k=0}^{\infty}\frac{\Gamma(\mu/2+1)}{k!\Gamma(\mu/2+1-k)}(d_{t,r}-a_i)^{2k}b_i^{\mu-2k}. \label{d-i-r}
\end{eqnarray}

We assume that $0 < a_i < d_{t,r}$ for each relay $i$ since the relays are interspersed throughout the region between the source and the destination.  Also, assume without loss of generality that $b_i > 0$ for each relay $i$ since $d_{t,i}$ and $d_{i,r}$ are functions of $b_i^2$.  Let the $m$ selected relays be located at $(a_1,b_1),(a_2,b_2),\ldots,(a_m,b_m)$.  Recall from \eqref{p-out-R1-m}, \eqref{p-out-R1-m-1}, \eqref{p-out-R1-m-4}, \eqref{p-out-R2-m}, \eqref{p-out-R2-m-1} and \eqref{p-out-R2-m-3} that in the high-SNR regime, $\bar{R}_{sc,2}(\mathcal{A})$ is a function of $G_{i,n}^{-2} = (d_{i,n})^{\mu}/\chi$.  Then, from \eqref{d-t-i} and \eqref{d-i-r}, we see that $\bar{R}_{sc,2}(\mathcal{A})$ is a function of $\{a_1,b_1,\ldots,a_m,b_m\}$.  Since we have assumed that $a_i$ and $b_i$ are positive for each relay $i$, and the binomial coefficients in \eqref{d-t-i} and \eqref{d-i-r} are not necessarily positive, $\bar{R}_{sc,2}(\mathcal{A})$ is a \textit{signomial function}\cite{BoyVan:ConvOpti:04} of $\{a_1,b_1,\ldots,a_m,b_m\}$ in the high-SNR regime.

$\textit{Signomial programs}$ usually do not admit efficient solutions via geometric programming unless the objective function and the associated inequality and equality constraints satisfy certain conditions \cite{BoyVan:ConvOpti:04}.  Next, we show that given a three-node line network, $\bar{R}_{sc,2}(\mathcal{A})$ reduces to a polynomial function of the relay position $d$.

\subsection{Optimal Relay Placement in Line Network}\label{three-nodes}
We consider a line network with $K_r = 1$.  The source is located at (0,0), the destination is located at $(d_{t,r},0)$ and the relay is located at $(d,0)$ where $0 < d < d_{t,r}$.  The outage probability $P_{out}(R_1,\mathcal{A})$ can be written as \cite{YukErk:BroaStraFadiRela:Oct:04}
\begin{eqnarray}
P_{out}(R_1,\mathcal{A}) & = & P(C_1(|h_{t,1}|^2) < R_1)P(C_1(|h_{t,r}|^2) < R_1) \nonumber \\
& + & P(C_1(|h_{t,1}|^2) \geq R_1, C_2(|h_{t,1}|^2) < R_2) \nonumber \\
& & \times P\bigg(\ln\bigg(1+\frac{|h_{t,r}|^2\beta P_t}{|h_{t,r}|^2\bar{\beta}P_t+\sigma^2}+\frac{|h_{1,r}|^2P_1}{\sigma^2}\bigg) < R_1\bigg) \nonumber \\
& + & P(C_2(|h_{t,1}|^2) \geq R_2)P\bigg(C_1\bigg(|h_{t,r}|^2+\frac{|h_{1,r}|^2P_1}{P_t}\bigg) < R_1\bigg) \label{p-out-R1}
\end{eqnarray}
and the outage probability $P_{out}(R_2,\mathcal{A})$ can be written as \cite{YukErk:BroaStraFadiRela:Oct:04}
\begin{eqnarray}
P_{out}(R_2,\mathcal{A}) & = & P(C_2(|h_{t,1}|^2) < R_2)P(C_2(|h_{t,r}|^2) < R_2) \nonumber \\
& + & P(C_2(|h_{t,1}|^2) \geq R_2)P\bigg(C_2\bigg(|h_{t,r}|^2+\frac{|h_{1,r}|^2P_1}{P_t}\bigg) < R_2\bigg). \label{p-out-R2}
\end{eqnarray}
As in Section \ref{m-nodes}, the expressions in \eqref{p-out-R1} and \eqref{p-out-R2} are fairly involved, so we again consider the high-SNR regime for ease of analysis.

In Appendix \ref{proof-thm-1}, we prove that \eqref{p-out-R1} simplifies to
\begin{eqnarray}
P_{out}(R_1,\mathcal{A}) & \sim & \Bigg(\frac{1}{G_{t,1}^2}\times\frac{\exp(R_1)-1}{(P_t/\sigma^2)\times(1-\bar{\beta}\exp(R_1))}\Bigg)\Bigg(\frac{1}{G_{t,r}^2}\times\frac{\exp(R_1)-1}{(P_t/\sigma^2)\times (1-\bar{\beta}\exp(R_1))}\Bigg) \nonumber \\
& + & \Bigg(\frac{1}{(P_1/P_t)G_{1,r}^2}\times\frac{\exp(R_1)-1}{(P_t/\sigma^2)\times (1-\bar{\beta}\exp(R_1))}\Bigg)\Bigg(\frac{1}{G_{t,r}^2}\times\frac{\exp(R_1)-1}{(P_t/\sigma^2)\times (1-\bar{\beta}\exp(R_1))}\Bigg)
\end{eqnarray}
and we prove that \eqref{p-out-R2} simplifies to
\begin{eqnarray}
P_{out}(R_2,\mathcal{A}) & \sim & \Bigg(\frac{1}{G_{t,1}^2}\times\frac{\exp(R_2)-1}{\bar{\beta}(P_t/\sigma^2)}\Bigg)\times\Bigg(\frac{1}{G_{t,r}^2}\times\frac{\exp(R_2)-1}{\bar{\beta}(P_t/\sigma^2)}\Bigg) \nonumber \\
& + & \frac{1}{2}\times\Bigg(\frac{1}{(P_1/P_t)G_{1,r}^2}\times\frac{\exp(R_2)-1}{\bar{\beta}(P_t/\sigma^2)}\Bigg)\times\Bigg(\frac{1}{G_{t,r}^2}\times\frac{\exp(R_2)-1}{\bar{\beta}(P_t/\sigma^2)}\Bigg).
\end{eqnarray}

Let $G_1 = (\exp(R_1)-1)/((P_t/\sigma^2)\times (1-\bar{\beta}\exp(R_1)))$ and $G_2 = (\exp(R_2)-1)/(\bar{\beta}(P_t/\sigma^2))$.  Then
\begin{eqnarray}
\bar{R}_{sc,2}(\mathcal{A}) & = & (1-P_{out}(R_1,\mathcal{A}))R_1 + (1-P_{out}(R_1,\mathcal{A}))(1-P_{out}(R_2,\mathcal{A}))R_2 \nonumber \\
& \sim & R_1\times (1-G_1^2\chi^2\times d_{t,r}^{\mu}(d^{\mu}+(P_t/P_1)\times(d_{t,r}-d)^{\mu})) \nonumber \\
& + & R_2\times (1-G_1^2\chi^2\times d_{t,r}^{\mu}(d^{\mu}+(P_t/P_1)\times(d_{t,r}-d)^{\mu})) \nonumber \\
& \times & (1-G_2^2\chi^2\times d_{t,r}^{\mu}(d^{\mu}+(1/2)\times(P_t/P_1)\times(d_{t,r}-d)^{\mu})).  \label{poly-opt}
\end{eqnarray}
For integral values of the path loss exponent $\mu$, finding the rate-maximizing relay position $\bar{d}$ is equivalent to maximizing a polynomial over $0 < d < d_{t,r}$.  For example, if $\mu = 2$, $\bar{R}_{sc,2}(\mathcal{A})$ is a fourth-degree polynomial in $d$.  Maximizing $\bar{R}_{sc,2}(\mathcal{A})$ with respect to $d$ is then equivalent to finding the roots of a cubic equation that lie in $0 < d < d_{t,r}$, assuming that at least one exists.

\section{Relay Subset Selection Algorithms}\label{subset-sel-alg}
The analysis in Section \ref{m-nodes} shows that the combinatorial optimization problem \eqref{signom-func} can be approximated by considering a given value of $m\in\{1,2,\ldots,K_r\}$ and maximizing a signomial function of the relay locations $(a_1,b_1),\ldots,(a_m,b_m)$ in the high-SNR regime, which yields the rate-maximizing set $(\bar{a}_1,\bar{b}_1),\ldots,(\bar{a}_m,\bar{b}_m)$.

Note that since $\bar{R}_{sc,2}(\mathcal{A})$ is a signomial function in the high-SNR regime, relays that are located close to any of the points in the rate-maximizing set should still yield high expected rates due to the inherent smoothness of signomial functions.  This motivates the following $\textit{proximity-based}$ algorithm for solving \eqref{signom-func}.

\newtheorem{opt-alg}{Algorithm}
\begin{opt-alg}
Multiple Fan Out
\end{opt-alg}
\begin{quote}
Step 1: For a given value of $m\in\{1,2,\ldots,K_r\}$, maximize $\bar{R}_{sc,2}(\mathcal{A})$ over all relay locations $(a_1,b_1),\ldots,(a_m,b_m)$ to find the rate-maximizing set $(\bar{a}_1,\bar{b}_1),\ldots,(\bar{a}_m,\bar{b}_m)$.

Step 2: Set $i = 1$ and $\mathcal{A} = \emptyset$.

Step 3: For relay $n$, where $1\leq n\leq K_r$, compute $d(n)$ where $d(n)$ is the distance from relay $n$ to $(\bar{a}_i,\bar{b}_i)$.  If relay $n$ is at location $(a_n,b_n)$, $d(n) = \sqrt{(a_n-\bar{a}_i)^2+(b_n-\bar{b}_i)^2}$.

Step 4: Find the closest relay $n$ to $(\bar{a}_i,\bar{b}_i)$ not in $\mathcal{A}$ and let $\mathcal{A} = \mathcal{A}\cup\{n\}$ and $P_n = P_{max}/m$.

Step 5: If $\|\mathcal{A}\| = m$, stop.  Otherwise, let $i = i+1$ and return to Step 3.
\end{quote}

We call the above relay selection algorithm $\textit{Multiple Fan Out}$ because the process of relay selection is analogous to a search party fanning out from its initial location.  Here, the objective is to ``fan out'' from $(\bar{a}_1,\bar{b}_1),\ldots,(\bar{a}_m,\bar{b}_m)$ until $m$ relays have been selected.

Note that in Step 2, we set $i = 1$ and increase $i$ in Step 5.  It turns out that $\bar{a}_1 = \cdots = \bar{a}_m$ and $\bar{b}_1 = \cdots = \bar{b}_m$ in Step 1, so the initial assignment of $i$ in Step 2 and its iteration in Step 5 are irrelevant.  The $\textit{Multiple Fan Out}$ algorithm, then, reduces to finding the $m$ closest relays to the single rate-maximizing point $(\bar{a}_1,\bar{b}_1)$.

Step 1 involves maximizing a signomial function, which usually does not admit an efficient solution.  To obtain a more tractable problem, the analysis in Section \ref{three-nodes} shows that in the case of a three-node line network, $\bar{R}_{sc,2}(\mathcal{A})$ is a polynomial function of the relay location $d$ in the high-SNR regime.  Maximizing $\bar{R}_{sc,2}(\mathcal{A})$ yields the rate-maximizing relay location $\bar{d}$.

Also, since $\bar{R}_{sc,2}(\mathcal{A})$ is a polynomial function in the high-SNR regime for a three-node line network, relays that are located close to the rate-maximizing relay location $\bar{d}$ should still yield high expected rates due to the inherent smoothness of polynomial functions.  This motivates another $\textit{proximity-based}$ algorithm for solving \eqref{signom-func}.  Again we assume that the path loss exponent $\mu$ takes on an integral value.  We also assume that exactly $m$ relays are to be selected, which further simplifies the algorithm.

\newtheorem{fan-out-alg}[opt-alg]{Algorithm}
\begin{fan-out-alg}
Single Fan Out
\end{fan-out-alg}
\begin{quote}
Step 1: Maximize \eqref{poly-opt} to find the rate-maximizing relay location $(\bar{d},0)$.

Step 2: For relay $n$, where $1\leq n\leq K_r$, compute $d(n)$ where $d(n)$ is the distance from relay $n$ to $(\bar{d},0)$.  If relay $n$ is at location $(a_n,b_n)$, $d(n) = \sqrt{(a_n-\bar{d})^2+b_n^2}$.

Step 3: Sort the set of relays $\{1,2,\ldots,K_r\}$ as $\{a_1,a_2,\ldots,a_{K_r}\}$, where \[d(a_1)\leq d(a_2)\leq\cdots\leq d(a_{K_r}).\]

Step 4: Find the closest relay $n$ to $(\bar{d},0)$ not in $\mathcal{A}$ and let $\mathcal{A} = \mathcal{A}\cup\{n\}$ and $P_n = P_{max}/m$.

Step 5: If $\|\mathcal{A}\| = m$, stop.  Otherwise, return to Step 4.
\end{quote}

We call the above relay selection algorithm $\textit{Single Fan Out}$ because $\bar{d}$ is computed via analysis of a single-relay line network.  Note that both the $\textit{Multiple Fan Out}$ and $\textit{Single Fan Out}$ algorithms are greedy strategies in that the order of procession through the list of relays is based on their proximity to $(\bar{x}_1,\bar{y}_1),\ldots,(\bar{x}_m,\bar{y}_m)$ and $\bar{d}$, respectively.  Greedy algorithms are useful for the problem at hand in that they possess an inherent simplicity, and their run-times are usually simple to characterize.  

We propose another greedy approach for selecting $\mathcal{A}$.  For simplicity, we assume that at most $m$ relays are to be selected.

\newtheorem{best-channel-gains}[opt-alg]{Algorithm}
\begin{best-channel-gains}
Best Gains
\end{best-channel-gains}
\begin{quote}
Step 1: For relay $i$, where $1\leq i\leq K_r$, compute $|h_{i,r}|^2$.

Step 2: Sort the set of relays $\{1,2,\ldots,K_r\}$ as $\{a_1,a_2,\ldots,a_{K_r}\}$, where \[|h_{a_1,r}|^2\geq |h_{a_2,r}|^2\geq\cdots\geq |h_{a_{K_r},r}|^2.\]

Step 3: Let $i = 1$ and $\mathcal{A} = \emptyset$.

Step 4: If relay $a_i$ has decoded $x_1$, then $\mathcal{A} = \mathcal{A}\cup\{a_i\}$ and $P_i = P_{max}/m$.

Step 5: If $\|\mathcal{A}\| = m$ or $i = K_r$, go to Step 6.  Otherwise, let $i = i+1$ and return to Step 4.

Step 6: If $\|\mathcal{A}\| < m$, let $P_i = P_{max}/\|\mathcal{A}\|$ for each relay $a_i\in\mathcal{A}$.
\end{quote}

We call the above relay selection algorithm $\textit{Best Gains}$ because the order of procession through the list of relays is based on their channel gains to the destination.  The objective is to choose relays that will be able to reliably transmit to the destination during time slot 2.  Note that a check is performed on each selected relay in Step 4 to ensure that it will be able to forward at least $x_1$ to the destination.

To obtain a lower bound on the performance of the above greedy algorithms, we propose the following algorithm whereby relays are randomly selected to transmit during time slot 2.

\newtheorem{random-relays}[opt-alg]{Algorithm}
\begin{random-relays}
Random Relays
\end{random-relays}
\begin{quote}
Step 1: Let $\mathcal{A} = \emptyset$.

Step 2: Randomly select a relay $i\in\{1,2,\ldots,K_r\}\setminus\mathcal{A}$ and let $\mathcal{A} = \mathcal{A}\cup\{i\}$ along with $P_i = P_{max}/m$.

Step 3: If $\|\mathcal{A}\| = m$, stop.  Otherwise, return to Step 2.
\end{quote}

Since the destination employs a diversity combining approach to receive the signals from all of the selected relays, the performance of the $\textit{Random Relays}$ algorithm should approach that of the other proposed relay selection algorithms as $m$ increases.

\section{Simulation Results}

\subsection{Performance of Relay Selection Algorithms}\label{perf-rel-sel-alg}
We place the source at $(0,0)$ and the destination at $(100,0)$.  We use the Worldwide Interoperability for Microwave Access (WiMAX) signaling bandwidth of 9 MHz \cite{WireMANWorkGrp}, and given a noise floor of -174dBm/Hz this yields a noise value $\sigma^2 = -104$dBm.  We also have a carrier frequency $f_c$ = 2.4GHz along with a reference distance $d_0$ = 1m and a path loss exponent $\mu$ = 3.  We randomly place $K_r = 20$ relays in the region between the source and the destination.

\begin{figure}[tb]
\begin{center}
\includegraphics[width=3.0in]{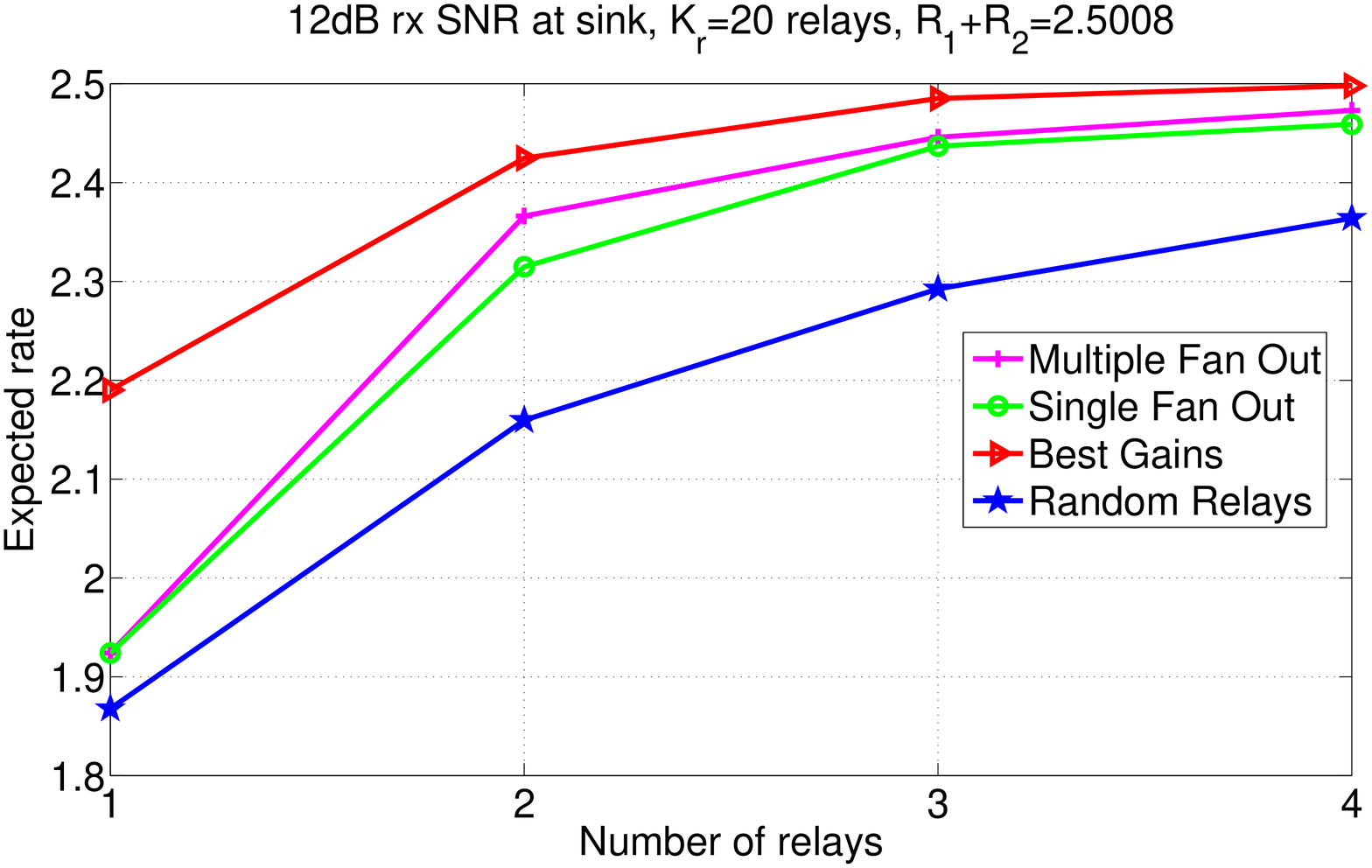}
\end{center}
\caption{Expected rate as a function of number of selected relays.}
\label{rate-relay-num}
\end{figure}

Fig. \ref{rate-relay-num} shows how the expected rate $\bar{R}_{sc,2}(\mathcal{A})$ varies with the number of selected relays $\|\mathcal{A}\| = m$ for the algorithms that we have proposed.  Here we fix the source's power $P_t = 6$dBm and the relay sum power constraint $P_{max} = P_t$.  The fraction of the source's power allocated to $x_1$ is $\beta = 0.75$ and the decoding thresholds for $x_1$ and $x_2$ are $|h_1| = 7.4\cdot 10^{-11}$ and $|h_2| = 1.25\cdot 10^{-10}$, respectively.  In the case of the Best Gains algorithm, we only consider cases where the number of selected relays $\|\mathcal{A}\| = m$.  We obtain the rate-maximizing set $(\bar{a}_1,\bar{b}_1),\ldots,(\bar{a}_m,\bar{b}_m)$ for the $\textit{Multiple Fan Out}$ algorithm via the $\texttt{fmincon}$ function from Matlab, which employs a sequential quadratic programming method.

We see that the greedy $\textit{Best Gains}$ algorithm yields the highest expected rate for all of the proposed selection strategies.  This is due to the fact that the $\textit{Best Gains}$ algorithm biases relay selection towards those relays that have good channel gains to the destination and can also transmit during time slot 2; this minimizes the chances of an outage event occurring at the destination where it cannot decode $x_1$.  On the other hand, the $\textit{Fan Out}$ algorithms select relays that are close to ergodic rate-maximizing points without considering their decoding status and their instantaneous channel gains to the destination.  Thus, the $\textit{Best Gains}$ algorithm attempts to optimize relay selection for each source transmission, though additional overhead is incurred relative to the $\textit{Fan Out}$ algorithms since the relays must inform the source of their decoding status and their channels to the destination.

Also, the $\textit{Single Fan Out}$ algorithm offers virtually the same performance as the $\textit{Multiple Fan Out}$ algorithm, which demonstrates the utility of our simplifications of the relay selection problem.  Here, the relays that are close to the rate-maximizing set for the $\textit{Multiple Fan Out}$ algorithm are also close to the rate-maximizing position $(\bar{d},0)$ for the $\textit{Single Fan Out}$ algorithm.  In addition, as the number of selected relays $\|\mathcal{A}\|$ increases, each strategy yields a higher expected rate which approaches the maximum expected rate.  Finally, note that the performance gap between all of the proposed strategies decreases as the number of selected relays increases.  This is due to the fact that selecting multiple relays yields an SNR gain at the destination that gradually overcomes the loss from selecting relays that might not be close to the rate-maximizing positions that are computed by the $\textit{Multiple Fan Out}$ and $\textit{Single Fan Out}$ algorithms.

\begin{figure}[tb]
\begin{center}
\includegraphics[width=3.0in]{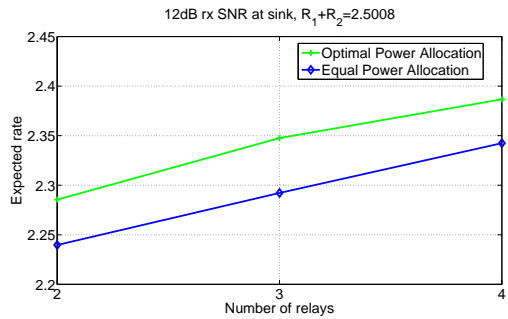}
\end{center}
\caption{Expected rate for two relay power allocation strategies.}
\label{rate-pwr-alloc}
\end{figure}

Fig. \ref{rate-pwr-alloc} shows how the expected rate $\bar{R}_{sc,2}(\mathcal{A})$ varies with the number of selected relays $\|\mathcal{A}\| = m$ for two relay power allocation strategies.  We use the same system parameters as in Fig. \ref{rate-relay-num}, except that we randomly place $m$ relays in the region between the source and the destination instead of $K_r = 20$ relays.  The Optimal Power Allocation strategy entails solving the relay selection problem in \eqref{signom-func}, and the Equal Power Allocation strategy assigns equal power to all of the selected relays.  We also set $P_{max} = P_t$.

We observe that the Equal Power Allocation strategy offers comparable performance to the Optimal Power Allocation strategy.  This illustrates the utility of low-complexity strategies that reduce the computation time inherent to interior-point methods that are needed to solve the optimization problem \eqref{signom-func}.

\begin{figure}[tb]
\begin{center}
\includegraphics[width=3.0in]{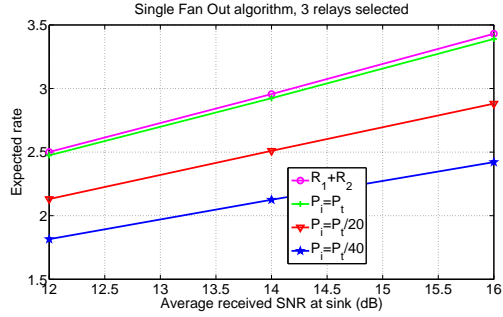}
\end{center}
\caption{Expected rate as a function of average received SNR at destination.}
\label{rate-snr-pwr}
\end{figure}

Fig. \ref{rate-snr-pwr} shows how the expected rate $\bar{R}_{sc,2}(\mathcal{A})$ varies with the average received SNR at the destination for different ratios between the relays' and source's powers.  When the average received SNR values at the destination are 0dB, 2dB and 4dB, the source's power takes on values $P_t = -6$dBm, $P_t = -4$dBm and $P_t = -2$dBm, respectively.

We see that as the average received SNR at the destination increases, the expected rate increases for each value of $P_i/P_t$.  Note that for a fixed value of $P_i/P_t$, increasing the average received SNR at the destination entails increasing $P_i$ and $P_t$.  For a fixed value of the average received SNR at the destination, the expected rate decreases as $P_i/P_t$ decreases, which corresponds to a decrease in $P_i$.  Thus, even though the three selected relays yield an SNR gain at the destination in time slot 2, this gain decreases as the relays' power decreases.

\begin{figure}[tb]
\begin{center}
\includegraphics[width=3.0in]{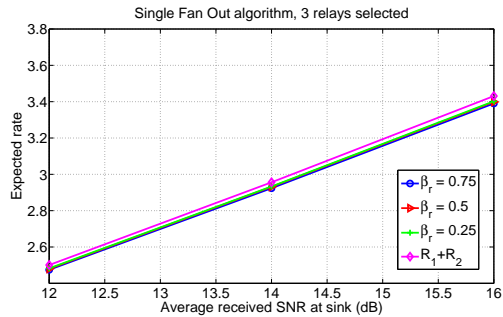}
\end{center}
\caption{Expected rate as a function of power split at relays.}
\label{rate-snr-beta}
\end{figure}

Fig. \ref{rate-snr-beta} shows how the expected rate $\bar{R}_{sc,2}(\mathcal{A})$ varies with the average received SNR at the destination for different values of the relays' power split $\beta_i$.  We set $\beta = 0.75$.

We see that as the relays' power split $\beta_i$ decreases, the expected rate increases for all average received SNR values at the destination.  Note that as $\beta_i$ decreases, $\bar{\beta_i}$ increases, which leads to an increase in $R_2$ as seen in \eqref{rate-x2}.  On the other hand, as $\bar{\beta_i}$ increases, \eqref{rate-x1} shows that $R_1$ decreases.  Fig. \ref{rate-snr-beta} shows that the increase in $R_2$ overcomes the decrease in $R_1$.

\section{Conclusion}
We have studied the problem of selecting a set of relay nodes to forward data in a two-hop wireless network.  We have considered a scenario where all relay nodes perform partial decode-and-forward operations based on a superposition coding strategy.  For this setup, we have shown that relay selection can be initially approximated by the problem of finding the relays that are close to a rate-maximizing location.  Finding the rate-maximizing location is usually computationally intensive, so we further simplify the relay selection problem by solving for the rate-maximizing location in a three-node line network.  These results motivate two $\textit{proximity-based}$ relay selection algorithms, where relays are chosen to forward data based on their proximity to one of the rate-maximizing locations.  We also demonstrated that the $\textit{proximity-based}$ algorithms outperform a random relay selection algorithm and yield rates close to those yielded by a greedy strategy that is based on channel state information.  In addition, we derived the diversity gain achieved by having multiple relays assist the source.  We also illustrated the performance impact of varying system parameters such as the ratio between the relays' and source's powers.

As noted in the Introduction, selecting the optimal subset of candidate relay nodes to assist a source is a difficult problem, and the proposed selection strategies are mainly intended to offer key insights.  In particular, the $\textit{proximity-based}$ algorithms motivate intelligent relay placement in a general two-hop static network with non-Rayleigh fading.  System designers can experiment with different network topologies and determine a throughput-maximizing configuration, where the achieved throughput would depend on the level of interference between the transmissions from distinct relays.  Also, the information-theoretic analysis in this paper can be modified to support more practical transmission strategies.  By applying cutting-edge coding strategies such as punctured low-density parity-check (LDPC) and turbo codes, the superposition coding approach that is employed in this paper can form the basis of a hybrid-ARQ strategy in a multihop network.

\appendix

\section{Proof of Theorem \ref{m-relays}}\label{proof-thm-1}
The probability that the destination cannot decode $x_1$ after time slot 2 is
\begin{eqnarray}
P_{out}(R_1,\mathcal{A}) & = & \sum_{(0\leq\alpha,\xi\leq m),\alpha+\xi\leq m}\Bigg(\sum_{\Delta\subseteq\mathcal{A},\Theta\subseteq\mathcal{A},\|\Delta\|=\alpha,\|\Theta\|=\xi,\Delta\bigcap\Theta=\emptyset}\Bigg(\prod_{\delta\in\Delta}P(C_1(|h_{t,\delta}|^2) < R_1)\Bigg) \nonumber \\
& & \times\Bigg(\prod_{\theta\in\Theta}P(C_1(|h_{t,\theta}|^2) \geq R_1, C_2(|h_{t,\theta}|^2) < R_2)\Bigg)\Bigg(\prod_{\eta\in (\mathcal{A}\setminus (\Delta\bigcup\Theta))}P(C_2(|h_{t,\eta}|^2) \geq R_2)\Bigg)\Bigg) \nonumber \\
& & \times P\Bigg(\ln\Bigg(1+\frac{|h_{t,r}|^2\beta P_t+\sum_{\eta\in (\mathcal{A}\setminus (\Delta\bigcup\Theta))}|h_{\eta,r}|^2\beta P_t}{|h_{t,r}|^2\bar{\beta}P_t+\sum_{\eta\in (\mathcal{A}\setminus (\Delta\bigcup\Theta))}|h_{\eta,r}|^2\bar{\beta} P_t+\sigma^2}+\sum_{\theta\in\Theta}\frac{|h_{\theta,r}|^2P_t}{\sigma^2}\Bigg) < R_1\Bigg)\Bigg). \label{p-out-R1-diversity-m}
\end{eqnarray}
Each term in the inner sum in \eqref{p-out-R1-diversity-m} represents a scenario where $\alpha$ selected relays cannot decode $x_1$, $\xi$ selected relays can decode $x_1$ but cannot decode $x_2$, and the remaining $m-\alpha-\xi$ selected relays can decode $x_2$.

Note that for a Rayleigh fading channel $h$,
\begin{eqnarray}
P(C_1(|h|^2) < R_1) & = & P\bigg(\ln\bigg(1+\frac{|h|^2\beta P_t}{|h|^2\bar{\beta}P_t+\sigma^2}\bigg) < R_1\bigg) \nonumber \\
& = & P\bigg(|h|^2 < \frac{\exp(R_1)-1}{1-\bar{\beta}\exp(R_1)}\times\frac{\sigma^2}{P_t}\bigg) \nonumber \\
& \sim & \frac{1}{\mathbb{E}(|h|^2)}\times\frac{\exp(R_1)-1}{(1-\bar{\beta}\exp(R_1))P_t/\sigma^2} \label{p-out-R1-diversity-1}
\end{eqnarray}
where \eqref{p-out-R1-diversity-1} follows from \cite[Fact 1]{LanTseETAL:CoopDiveWireNetw:Dec:04}.

Also, for a Rayleigh fading channel $h$, 
\begin{eqnarray}
P(C_1(|h|^2) \geq R_1, C_2(|h|^2) < R_2)& \leq & P(C_1(|h|^2) \geq R_1) \nonumber \\
& \sim & 1. \label{p-out-R1-diversity-2}
\end{eqnarray}

In addition, for independent Rayleigh fading channels $h_1$ and $h_2$, note that
\begin{eqnarray}
P\bigg(\ln\bigg(1+\frac{|h_1|^2\beta P_t}{|h_1|^2\bar{\beta}P_t+\sigma^2}+\frac{|h_2|^2P_t}{\sigma^2}\bigg) < R_1\bigg) & \leq & P\bigg(\ln\bigg(1+\frac{|h_1|^2\beta P_t+|h_2|^2\beta P_t}{|h_1|^2\bar{\beta}P_t+|h_2|^2\bar{\beta}P_t+\sigma^2}\bigg) < R_1\bigg) \nonumber \\
& = & P(C_1(|h_1|^2 + |h_2|^2) < R_1) \nonumber \\
& = & P\bigg(|h_1|^2 + |h_2|^2 < \frac{\exp(R_1)-1}{1-\bar{\beta}\exp(R_1)}\times\frac{\sigma^2}{P_t}\bigg) \nonumber \\
& \sim & \frac{1}{2\mathbb{E}(|h_1|^2)\mathbb{E}(|h_2|^2)}\bigg(\frac{\exp(R_1)-1}{(1-\bar{\beta}\exp(R_1))P_t/\sigma^2}\bigg)^2 \label{p-out-R1-diversity-3}
\end{eqnarray}
where \eqref{p-out-R1-diversity-3} follows from \cite[Fact 2]{LanTseETAL:CoopDiveWireNetw:Dec:04}.

Also, for a Rayleigh fading channel $h$, 
\begin{equation}
P(C_2(|h|^2) \geq R_2)\sim 1.  \label{p-out-R1-diversity-4}
\end{equation}

In addition, for independent Rayleigh fading channels $h_1$, $h_2$ and $h_3$, note that
\begin{equation}
\begin{array}{lll}
P\bigg(\ln\bigg(1+\frac{|h_1|^2\beta P_t}{|h_1|^2\bar{\beta}P_t+\sigma^2}+\frac{|h_2|^2P_t}{\sigma^2}+\frac{|h_3|^2P_t}{\sigma^2}\bigg) < R_1\bigg) & \leq & P\bigg(\ln\bigg(1+\frac{|h_1|^2\beta P_t+|h_2|^2\beta P_t+|h_3|^2\beta P_t}{|h_1|^2\bar{\beta}P_t+|h_2|^2\bar{\beta}P_t+|h_3|^2\bar{\beta}P_t+\sigma^2}\bigg) < R_1\bigg) \\
& = & P(C_1(|h_1|^2 + |h_2|^2 + |h_3|^2) < R_1) \\
& = & P\bigg(|h_1|^2 + |h_2|^2 + |h_3|^2< \frac{\exp(R_1)-1}{1-\bar{\beta}\exp(R_1)}\times\frac{\sigma^2}{P_t}\bigg) \\
& \sim & \frac{1}{6\mathbb{E}(|h_1|^2)\mathbb{E}(|h_2|^2)\mathbb{E}(|h_3|^2)}\Big(\frac{\exp(R_1)-1}{(1-\bar{\beta}\exp(R_1))P_t/\sigma^2}\Big)^3 \label{p-out-R1-diversity-5}
\end{array}
\end{equation}
where \eqref{p-out-R1-diversity-5} follows from \cite[Appendix B]{LanWor:DistSpacTimeCode:Oct:03}.

We use \eqref{p-out-R1-diversity-1}, \eqref{p-out-R1-diversity-2}, \eqref{p-out-R1-diversity-3}, \eqref{p-out-R1-diversity-4} and \eqref{p-out-R1-diversity-5} to see that
\begin{eqnarray}
& & P\Bigg(\ln\Bigg(1+\frac{|h_{t,r}|^2\beta P_t+\sum_{\eta\in (\mathcal{A}\setminus (\Delta\bigcup\Theta))}|h_{\eta,r}|^2\beta P_t}{|h_{t,r}|^2\bar{\beta}P_t+\sum_{\eta\in (\mathcal{A}\setminus (\Delta\bigcup\Theta))}|h_{\eta,r}|^2\bar{\beta} P_t+\sigma^2}+\sum_{\theta\in\Theta}\frac{|h_{\theta,r}|^2P_t}{\sigma^2}\Bigg) < R_1\Bigg) \label{p-out-R1-diversity} \\
& \leq & P\Bigg(\ln\Bigg(1+\frac{|h_{t,r}|^2\beta P_t+\sum_{\nu\in (\mathcal{A}\setminus\Delta)}|h_{\nu,r}|^2\beta P_t}{|h_{t,r}|^2\bar{\beta}P_t+\sum_{\nu\in (\mathcal{A}\setminus\Delta)}|h_{\nu,r}|^2\bar{\beta} P_t+\sigma^2}\Bigg) < R_1\Bigg) \nonumber \\
& = & P\Bigg(C_1\Bigg(|h_{t,r}|^2 + \sum_{\nu\in (\mathcal{A}\setminus\Delta)}|h_{\nu,r}|^2\Bigg) < R_1\Bigg) \nonumber \\
& \sim & \frac{1}{(m-\alpha + 1)!}\times\frac{1}{\mathbb{E}(|h_{t,r}|^2)}\times\bigg(\frac{\exp(R_1)-1}{(1-\bar{\beta}\exp(R_1))P_t/\sigma^2}\bigg)^{-(m-\alpha + 1)}\prod_{\nu\in (\mathcal{A}\setminus\Delta)}\frac{1}{\mathbb{E}(|h_{\nu,r}|^2)}. \nonumber
\end{eqnarray}

Thus, the high-SNR behavior of $P_{out}(R_1,\mathcal{A})$ is
\begin{eqnarray}
P_{out}(R_1,\mathcal{A}) & \sim & \Bigg(\Bigg(\frac{P_t}{\sigma^2}\Bigg)^{-1}\Bigg)^{\alpha}\times (1)^{\beta}\times (1)^{m-\alpha-\beta} \times \Bigg(\frac{P_t}{\sigma^2}\Bigg)^{-(m-\alpha + 1)} \nonumber \\
& = & \Bigg(\frac{P_t}{\sigma^2}\Bigg)^{-(m+1)}
\end{eqnarray}
and so we obtain a diversity gain of $\kappa_1(m) = m+1$ for decoding $x_1$ at the destination.

The probability that the destination cannot decode $x_2$ after time slot 2 is
\begin{eqnarray}
P_{out}(R_2,\mathcal{A}) & = & \sum_{0\leq\alpha\leq m}\Bigg(\sum_{\|\Delta\|=\alpha,\Delta\subseteq\mathcal{A}}\Bigg(\prod_{\delta\in\Delta}P(C_2(|h_{t,\delta}|^2) < R_2)\Bigg) \nonumber \\
& & \times\Bigg(\prod_{\theta\in (\mathcal{A}\setminus\Delta)}P(C_2(|h_{t,\theta}|^2) \geq R_2)\Bigg) \nonumber \\
& & \times P\Bigg(C_2\Bigg(|h_{t,r}|^2+\sum_{\theta\in (\mathcal{A}\setminus\Delta)}|h_{\theta,r}|^2\Bigg) < R_2\Bigg)\Bigg). \label{p-out-R2-diversity-m}
\end{eqnarray}
Each term in the inner sum in \eqref{p-out-R2-diversity-m} represents a decoding scenario where $\alpha$ selected relays cannot decode $x_2$ and the remaining $m-\alpha$ selected relays can decode $x_2$.

Note that for a Rayleigh fading channel $h$,
\begin{eqnarray}
P(C_2(|h|^2) < R_2) & = & P\bigg(\ln\bigg(1+\frac{|h|^2\bar{\beta}P_t}{\sigma^2}\bigg) < R_2\bigg) \nonumber \\
& = & P\bigg(|h|^2 < \frac{\exp(R_2)-1}{\bar{\beta}}\times\frac{\sigma^2}{P_t}\bigg) \nonumber \\
& \sim & \frac{1}{\mathbb{E}(|h|^2)}\times\frac{\exp(R_2)-1}{\bar{\beta}P_t/\sigma^2} \label{p-out-R2-diversity-1}
\end{eqnarray}
where \eqref{p-out-R2-diversity-1} follows from \cite[Fact 1]{LanTseETAL:CoopDiveWireNetw:Dec:04}.

Also, for independent Rayleigh fading channels $h_1$ and $h_2$, note that
\begin{eqnarray}
P(C_2(|h_1|^2 + |h_2|^2) < R_2) & = & P\bigg(\ln\bigg(1+\frac{|h_1|^2\bar{\beta}P_t}{\sigma^2}+\frac{|h_2|^2\bar{\beta}P_t}{\sigma^2}\bigg) < R_2\bigg) \nonumber \\
& = & P\bigg(|h_1|^2 + |h_2|^2 < \frac{\exp(R_2)-1}{\bar{\beta}}\times\frac{\sigma^2}{P_t}\bigg) \nonumber \\
& \sim & \frac{1}{2\mathbb{E}(|h_1|^2)\mathbb{E}(|h_2|^2)}\bigg(\frac{\exp(R_2)-1}{\bar{\beta}P_t/\sigma^2}\bigg)^2 \label{p-out-R2-diversity-2}
\end{eqnarray}
where \eqref{p-out-R2-diversity-2} follows from \cite[Fact 2]{LanTseETAL:CoopDiveWireNetw:Dec:04}.

In addition, for independent Rayleigh fading channels $h_1$, $h_2$ and $h_3$, note that
\begin{eqnarray}
P(C_2(|h_1|^2 + |h_2|^2 + |h_3|^2) < R_2) & = & P\bigg(\ln\bigg(1+\frac{|h_1|^2\bar{\beta}P_t}{\sigma^2}+\frac{|h_2|^2\bar{\beta}P_t}{\sigma^2}+\frac{|h_3|^2\bar{\beta}P_t}{\sigma^2}\bigg) < R_2\bigg) \nonumber \\
& = & P\bigg(|h_1|^2 + |h_2|^2 + |h_3|^2 < \frac{\exp(R_2)-1}{\bar{\beta}}\times\frac{\sigma^2}{P_t}\bigg) \nonumber \\
& \sim & \frac{1}{6\mathbb{E}(|h_1|^2)\mathbb{E}(|h_2|^2)\mathbb{E}(|h_3|^2)}\bigg(\frac{\exp(R_2)-1}{\bar{\beta}P_t/\sigma^2}\bigg)^3 \label{p-out-R2-diversity-3}
\end{eqnarray}
where \eqref{p-out-R2-diversity-3} follows from \cite[Appendix B]{LanWor:DistSpacTimeCode:Oct:03}.

We use \eqref{p-out-R1-diversity-4}, \eqref{p-out-R2-diversity-1}, \eqref{p-out-R2-diversity-2} and \eqref{p-out-R2-diversity-3} to see that 
\begin{eqnarray}
P\Bigg(C_2\Bigg(|h_{t,r}|^2+\sum_{\theta\in (\mathcal{A}\setminus\Delta)}|h_{\theta,r}|^2\Bigg) < R_2\Bigg) & & \label{p-out-R2-diversity} \\ \nonumber 
\sim \frac{1}{(m-\alpha + 1)!}\times\frac{1}{\mathbb{E}(|h_{t,r}|^2)}\times\bigg(\frac{\exp(R_2)-1}{\bar{\beta}P_t/\sigma^2}\bigg)^{-(m-\alpha + 1)}\prod_{\nu\in (\mathcal{A}\setminus\Delta)}\frac{1}{\mathbb{E}(|h_{\nu,r}|^2)}. & &
\end{eqnarray}

Thus, the high-SNR behavior of $P_{out}(R_2,\mathcal{A})$ is
\begin{eqnarray}
P_{out}(R_2,\mathcal{A}) & \sim & \Bigg(\Bigg(\frac{P_t}{\sigma^2}\Bigg)^{-1}\Bigg)^{\alpha}\times (1)^{m-\alpha} \times \Bigg(\frac{P_t}{\sigma^2}\Bigg)^{-(m-\alpha + 1)} \nonumber \\
& = & \Bigg(\frac{P_t}{\sigma^2}\Bigg)^{-(m+1)}
\end{eqnarray}
and so we obtain a diversity gain of $\kappa_2(m) = m+1$ for decoding $x_2$ at the destination.

Thus, we conclude that selecting $m$ relays allows us to reap a diversity gain of $m+1$ for both $R_1$ and $R_2$.

\section{Proof of Theorem \ref{m-relays-general}}\label{proof-thm-2}
The proof of Theorem \ref{m-relays-general} is similar to that of Theorem \ref{m-relays} in Appendix \ref{proof-thm-1}.

First, we consider the decoding of $x_1$ at the destination.  Recalling that $(P_i/\sigma^2) = (P_t/\sigma^2)^k$ for each relay $i$, we can use \eqref{p-out-R1-diversity} to see that
\begin{eqnarray}
& & P\Bigg(\ln\Bigg(1+\frac{|h_{t,r}|^2\beta P_t+\sum_{\eta\in (\mathcal{A}\setminus (\Delta\bigcup\Theta))}|h_{\eta,r}|^2\beta P_{\eta}}{|h_{t,r}|^2\bar{\beta}P_t+\sum_{\eta\in (\mathcal{A}\setminus (\Delta\bigcup\Theta))}|h_{\eta,r}|^2\bar{\beta} P_{\eta}+\sigma^2}+\sum_{\theta\in\Theta}\frac{|h_{\theta,r}|^2P_{\theta}}{\sigma^2}\Bigg) < R_1\Bigg) \label{p-out-R1-general} \\
& \leq & P\Bigg(\ln\Bigg(1+\frac{|h_{t,r}|^2\beta P_t+\sum_{\nu\in (\mathcal{A}\setminus\Delta)}|h_{\nu,r}|^2\beta P_{\nu}}{|h_{t,r}|^2\bar{\beta}P_t+\sum_{\nu\in (\mathcal{A}\setminus\Delta)}|h_{\nu,r}|^2\bar{\beta} P_{\nu}+\sigma^2}\Bigg) < R_1\Bigg) \nonumber \\
& \sim & \frac{1}{(m-\alpha + 1)!}\times\frac{1}{\mathbb{E}((P_t/\sigma^2)|h_{t,r}|^2)}\times\bigg(\frac{\exp(R_1)-1}{(1-\bar{\beta}\exp(R_1))}\bigg)^{-(m-\alpha + 1)}\prod_{\nu\in (\mathcal{A}\setminus\Delta)}\frac{1}{\mathbb{E}((P_{\nu}/\sigma^2)|h_{\nu,r}|^2)} \nonumber \\
& = & \frac{1}{(m-\alpha + 1)!}\times\frac{1}{\mathbb{E}(|h_{t,r}|^2)}\times\bigg(\frac{\exp(R_1)-1}{(1-\bar{\beta}\exp(R_1))}\bigg)^{-(m-\alpha + 1)}\Bigg(\frac{P_t}{\sigma^2}\Bigg)^{-(k(m-\alpha)+1)}\prod_{\nu\in (\mathcal{A}\setminus\Delta)}\frac{1}{\mathbb{E}(|h_{\nu,r}|^2)}. \nonumber
\end{eqnarray}

Thus, for a given integer value of $\alpha\in\{0,\ldots,m\}$, the high-SNR behavior of $P_{out}(R_1,\mathcal{A})$ is
\begin{eqnarray}
P_{out}(R_1,\mathcal{A}) & \sim & \Bigg(\Bigg(\frac{P_t}{\sigma^2}\Bigg)^{-1}\Bigg)^{\alpha}\times (1)^{\beta}\times (1)^{m-\alpha-\beta} \times \Bigg(\frac{P_t}{\sigma^2}\Bigg)^{-(k(m-\alpha)+1)} \nonumber \\
& = & \Bigg(\frac{P_t}{\sigma^2}\Bigg)^{-(km+1+\alpha(1-k))}.
\end{eqnarray}
We then minimize $km+1+\alpha(1-k)$ over all $\alpha\in\{0,\ldots,m\}$ to obtain the generalized diversity gain $\kappa_1^g(m)$ in Theorem \ref{m-relays-general}.

We then consider the decoding of $x_2$ at the destination.  Recalling that $(P_i/\sigma^2) = (P_t/\sigma^2)^k$ for each relay $i$, we can use \eqref{p-out-R2-diversity} to see that
\begin{eqnarray}
P\bigg(\ln\bigg(1+\frac{|h_{t,r}|^2\bar{\beta}P_t}{\sigma^2}+\sum_{\theta\in (\mathcal{A}\setminus\Delta)}\frac{|h_{\theta,r}|^2\bar{\beta}P_{\theta}}{\sigma^2}\bigg) < R_2\bigg) \nonumber \\
\sim \frac{1}{(m-\alpha + 1)!}\times\frac{1}{\mathbb{E}((P_t/\sigma^2)|h_{t,r}|^2)}\times\bigg(\frac{\exp(R_2)-1}{\bar{\beta}}\bigg)^{-(m-\alpha + 1)}\prod_{\nu\in (\mathcal{A}\setminus\Delta)}\frac{1}{\mathbb{E}((P_{\nu}/\sigma^2)|h_{\nu,r}|^2)} & & \nonumber \\
= \frac{1}{(m-\alpha + 1)!}\times\frac{1}{\mathbb{E}(|h_{t,r}|^2)}\times\bigg(\frac{\exp(R_2)-1}{\bar{\beta}}\bigg)^{-(m-\alpha + 1)}\Bigg(\frac{P_t}{\sigma^2}\Bigg)^{-(k(m-\alpha)+1)}\prod_{\nu\in (\mathcal{A}\setminus\Delta)}\frac{1}{\mathbb{E}(|h_{\nu,r}|^2)}. & & \nonumber
\end{eqnarray}

Thus, for a given integer value of $\alpha\in\{0,\ldots,m\}$, the high-SNR behavior of $P_{out}(R_2,\mathcal{A})$ is
\begin{eqnarray}
P_{out}(R_2,\mathcal{A}) & \sim & \Bigg(\Bigg(\frac{P_t}{\sigma^2}\Bigg)^{-1}\Bigg)^{\alpha}\times (1)^{m-\alpha} \times \Bigg(\frac{P_t}{\sigma^2}\Bigg)^{-k(m-\alpha)+1)} \nonumber \\
& = & \Bigg(\frac{P_t}{\sigma^2}\Bigg)^{-(km+1+\alpha(1-k))}.
\end{eqnarray}
We then minimize $km+1+\alpha(1-k)$ over all $\alpha\in\{0,\ldots,m\}$ to obtain the generalized diversity gain $\kappa_2^g(m) = \kappa_1^g(m)$ in Theorem \ref{m-relays-general}.

\end{document}